\documentclass[apj]{emulateapj}
\usepackage{apjfonts}

\begin{document}

\title{Very-High-Energy $\gamma$-Ray Observations of a Strong Flaring Activity in M\,87 in 2008 February}
\shorttitle{Strong Flaring Activity in M\,87 in 2008 February}
\shortauthors{The MAGIC Collaboration}

%
\author{
J.~Albert\altaffilmark{a},
E.~Aliu\altaffilmark{b},
H.~Anderhub\altaffilmark{c},
L.~A.~Antonelli\altaffilmark{d},
P.~Antoranz\altaffilmark{e},
M.~Backes\altaffilmark{f},
C.~Baixeras\altaffilmark{g},
J.~A.~Barrio\altaffilmark{e},
H.~Bartko\altaffilmark{h},
D.~Bastieri\altaffilmark{i}
% ...
}
\altaffiltext{a} {Universit\"at W\"urzburg, D-97074 W\"urzburg, Germany}
\altaffiltext{b} {IFAE, Edifici Cn., Campus UAB, E-08193 Bellaterra, Spain}
\altaffiltext{c} {ETH Zurich, CH-8093 Switzerland}
\altaffiltext{d} {INAF National Institute for Astrophysics, I-00136 Rome, Italy}
\altaffiltext{e} {Universidad Complutense, E-28040 Madrid, Spain}
\altaffiltext{f} {Technische Universit\"at Dortmund, D-44221 Dortmund, Germany}
\altaffiltext{g} {Universitat Aut\`onoma de Barcelona, E-08193 Bellaterra, Spain}
\altaffiltext{h} {Max-Planck-Institut f\"ur Physik, D-80805 M\"unchen, Germany}
\altaffiltext{i} {Universit\`a di Padova and INFN, I-35131 Padova, Italy}
% ...

\begin{abstract}
M\,87 is the only known non blazar radio galaxy to emit very high
energy (VHE) gamma-rays. During a monitoring program of M\,87, a
rapid flare in VHE gamma-rays was detected by the MAGIC telescope
in early 2008. The flux was found to be variable above 350\,GeV on
a timescale as short as 1 day at a significance level of
$5.6\sigma$. The highest measured flux reached 15\% of the Crab
Nebula flux. We observed several substantial changes of the flux
level during the 13 day observing period. The flux at lower
energies (150 -- 350\,GeV), instead, is compatible with being
constant. The energy spectrum can be described by a power law with
a photon index of $2.30 \pm 0.11_\mathrm{stat} \pm
0.20_\mathrm{syst}$. The observed day-scale flux variability at
VHE prefers the M\,87 core as source of the emission and implies
that either the emission region is very compact (just a few
Schwarzschild radii) or the Doppler factor of the emitting blob is
rather large in the case of a non expanding emission region.
\end{abstract}


\section{Introduction}
M\,87 is a giant elliptical radio galaxy (RG) of
Fanaroff-Riley-I-type \citep[FR\,I,][]{FanaroffRiley74} in the
Virgo Cluster at a distance of 16\,Mpc \citep{Macri99}. It is
powered by a supermassive black hole (BH) of
$(3.2\pm0.9)\times10^{9} \mathrm{M}_\odot$ \citep{Macchetto97}.
The M\,87 jet was the first-ever observed \citep{Curtis18}, and
due to the proximity of M\,87, its morphological substructures can
be resolved and a unique view of its innermost regions is
possible. The jet, originating from the RG core, extends to 20''
(\citealt{Marshall02}; equivalent to a 2\,kpc projected linear
distance). Several compact regions (``knots'') along its axis are
resolved in the radio, optical, and X-ray regimes.
These knots have similar morphologies in all
wave bands, although the X-ray knots appear to be tens of pc closer
to the core than the optical and radio knots \citep{WilsonYang02}.
The variable brightness of the knots may be due to several shock
fronts in the jet, being responsible for particle acceleration and
non-thermal emission. Superluminal motion of the knots has been
observed in the optical \citep{Biretta96} and radio
\citep{Forman07} wave bands, constraining the viewing
angle of the jet to $<43^{\circ}\pm4^{\circ}$.

The innermost resolved bright knot HST-1 is located at 0.82''
(64\,pc) from the core and is the most prominent feature of the
jet. HST-1 has shown many flares exceeding the luminosity of the
M\,87 core emission. Its X-ray brightness has increased by a
factor $>50$ from 2000 to 2005 \citep{Harris06}. A 
correlation between radio, optical, and X-ray luminosity points to
a common origin of the emission. The measured superluminal motion
in HST-1 is higher than in other knots, suggesting a viewing angle
of $<19^{\circ}$ for this part of the jet. The core itself is
variable, too, and also shows a correlation between the
emission levels from radio frequencies through X-rays
\citep{Perlman03}.

M\,87 was not detected by EGRET. The first hint of
very high energy (VHE; $>250$\,GeV) $\gamma$-ray emission was
reported by \citet{HEGRA-M87}, and later
confirmed by \citet{HESS-M87} and 
\citet{VERITAS-M87}. The emission is variable on a
timescale of years. 
The reported $\approx 2$-day variability \citep{HESS-M87} narrows down
the size of the emission region to be on the order of the light-crossing
time of the central BH.
With its expected low accretion rate, the M\,87
core radiation is not strong enough to attenuate significantly TeV
$\gamma$-rays even at 5 Schwarzschild radii ($R_S$) away from
the BH \citep{na07}. All this implies a production region in the
immediate vicinity of the M\,87 core. 
During later observations, no significant
flux variation was found \citep{VERITAS-M87}.
An X-ray--VHE $\gamma$-ray correlation is expected in most emission models, but
was not unambiguously found so far. Whereas \citet{HESS-M87} claim
a hint of a correlation between the soft (0.3--10 keV) X-rays at
HST-1 and the VHE $\gamma$-rays, \citet{VERITAS-M87} find a
year-by-year correlation between the 2--10 keV X-ray flux of the
M\,87 core and the VHE $\gamma$-ray emission instead, but do not
observe a correlation between the two energy bands on shorter
timescales.

The radio to X-ray emission of the jet is due to non-thermal
synchrotron radiation of relativistic electrons in the jet. The
observed knots and flares in M\,87 point to a complicated
morphology with several shock fronts producing these electrons.
While all 23 currently known extragalactic VHE $\gamma$-ray
emitters\footnote{See, e.g.,
http://www.mppmu.mpg.de/$\sim$rwagner/sources/ for an up-to-date
list} are blazars, M\,87 is assumed to be a blazar not aligned to
our line of sight \citep{Tsv98}. If the observed VHE emission from
M\,87 is associated with the innermost part of its jet, then
blazar emission models may hold. In blazars without prominent disk
or broad-line features, the VHE emission is explained by inverse
Compton processes involving the synchrotron photons and their
parent electron population \citep[synchrotron self-Compton models,
e.g.][]{Maraschi92}. Alternatively, in hadronic models,
interactions of a highly relativistic jet outflow with ambient
matter \citep{DarLaor,Beall99}, proton-induced cascades
\citep{PIC}, or synchrotron proton radiation
\citep{Muecke01,Aharonian00} may produce VHE $\gamma$-rays. In such a
scenario, M\,87 might also account for parts of the observed
ultra-high-energy cosmic rays \citep{uhecr}. It should be noted,
that for M\,87 the location of the VHE emission is still
uncertain. Specific emission models for high-energy processes
close to the core \citep{gpk05,ghi05,len07,tg08}, in the
large-scale jet \citep{sta03,hon07}, or in the vicinity of a BH
\citep{na07,riegeraha08} have been developed.
%

The MAGIC collaboration
performed monitoring observations of M\,87 starting from 2008 January, 
sharing the task with the VERITAS and H.E.S.S.\ experiments.
Here we report on MAGIC results from a subset of these data,
revealing a variability timescale of M\,87 of 1 day. The energy
spectrum and light curves are discussed.


\section{Observations and Data Analysis}
The MAGIC telescope is located on the Canary Island of La Palma
(2200\,m above sea level, 28$^\circ$45$'$N, 17$^\circ$54$'$W).
MAGIC is a stand-alone imaging air Cerenkov telescope (IACT) with
a 17~m-diameter tessellated reflector dish.
MAGIC has a low energy threshold of $50-60$
GeV (trigger threshold at small zenith angles). The accessible
energy range extends up to tens of TeV with a typical energy
resolution of 20\%--30\%, depending on the zenith angle and energy
\citep{crab}.

The data set comprises observations from 2008 January 30 to 2008
February 11. These were performed in the wobble mode
\citep{daum} for 26.7~hr. The zenith angle of the observations
ranges from 16$^{\circ}$ to 35$^{\circ}$. After removing runs with unusually
low trigger rates, mostly caused by bad weather conditions, the
effective observing time amounts to 22.8~hr.

The data were analyzed using the MAGIC standard calibration and
analysis \citep{crab}. The analysis is based on image parameters
\citep{hillas,tescaro07} and the random forest \citep[RF,][]{magicrf} method,
which are used to define the so-called hadronness of each event.
The cut in hadronness for $\gamma$/hadron separation was optimized on a
contemporaneous data set of the Crab Nebula. After this cut the
distribution of the angle $ALPHA$, which is the angle between the
main image axis and the line between center of gravity of the
image and the source position in the camera, is used to determine
the signal in the ON-source region. Three background (OFF) sky
regions are chosen symmetrically to the ON-source region with
respect to the camera center. The final cut $|ALPHA|<5^{\circ}$
(Fig.~\ref{fig:alpha}) was also optimized on the Crab Nebula data
to determine the number of excess events and the significance of the signal.

The energies of the $\gamma$-ray candidates were also estimated
using the RF method. To derive a differential energy
spectrum, we applied looser cuts than those in
Fig.~\ref{fig:alpha} to retain a higher number of $\gamma$-ray
candidates and to lower the effective analysis threshold down to
150\,GeV. Looser cuts also reduce systematic uncertainties between
data and Monte-Carlo events, which is important for the estimation
of the effective collection areas. The derived spectrum was
unfolded to correct for the effects of the limited energy
resolution of the detector \citep{magicuf}. Finally, the spectrum
and the light curves were corrected for trigger inefficiencies due
to higher discriminator thresholds during partial moon light and
twilight conditions \citep{magicmoon}. These corrections are on
the order of 0\%--20\%.
The data were also
analyzed with an independent analysis yielding, within statistical
errors, the same results. 

%
\section{Results}
%
%
\begin{figure}
\begin{center}
\includegraphics[width=0.875\linewidth]{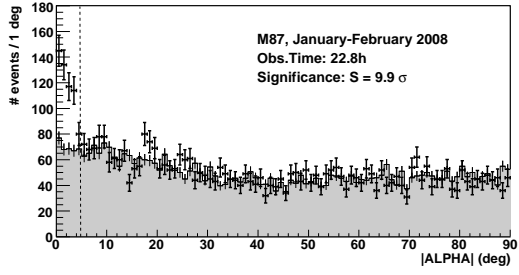}
\caption{\label{fig:alpha} $|ALPHA|$ distribution for the overall data sample.
The background (gray histogram) is estimated using three OFF
regions arranged symmetrically to the ON-source region with
respect to the camera center. A $\gamma$-ray excess with a
significance of 9.9 standard deviations is obtained.}
\end{center}
\end{figure}

The $|ALPHA|$ distribution after so-called tight cuts is shown in
Fig.~\ref{fig:alpha}. The applied cuts are $SIZE > 450$\,photoelectrons
and hadronness $h < 0.02$.
After the final $|ALPHA|$ cut (resulting in an overall cut
efficiency of 37\%), the total signal of 241 events over 349 normalized
background events corresponds to an excess with a significance of
$9.9\sigma$ along eq. 17 in \citet{LiMa}.
The highest flux was observed on
2008 February 1 at a significance of $8.0\sigma$.

%
%
\begin{figure}
\begin{center}
\includegraphics[width=0.875\linewidth]{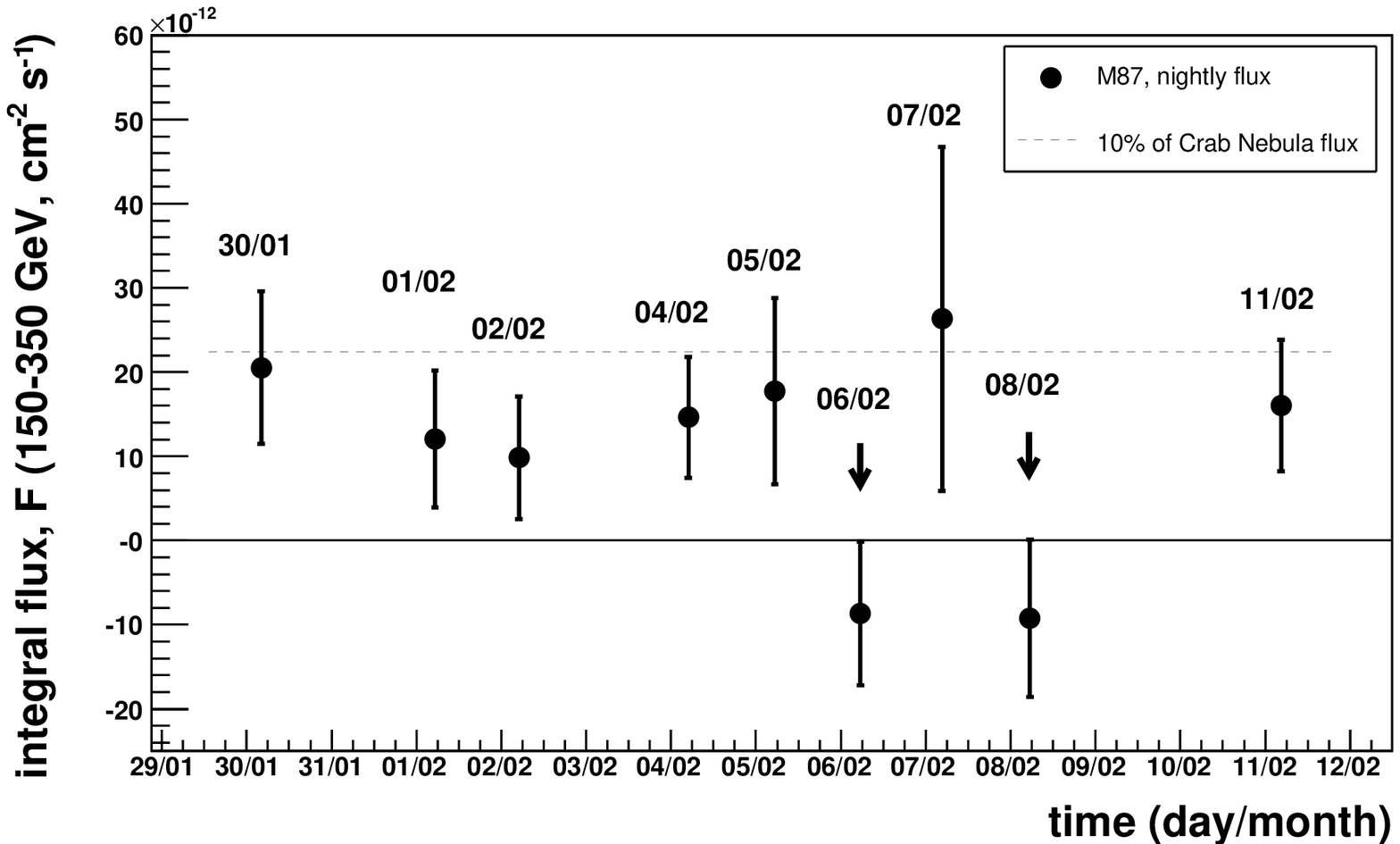}
\includegraphics[width=0.875\linewidth]{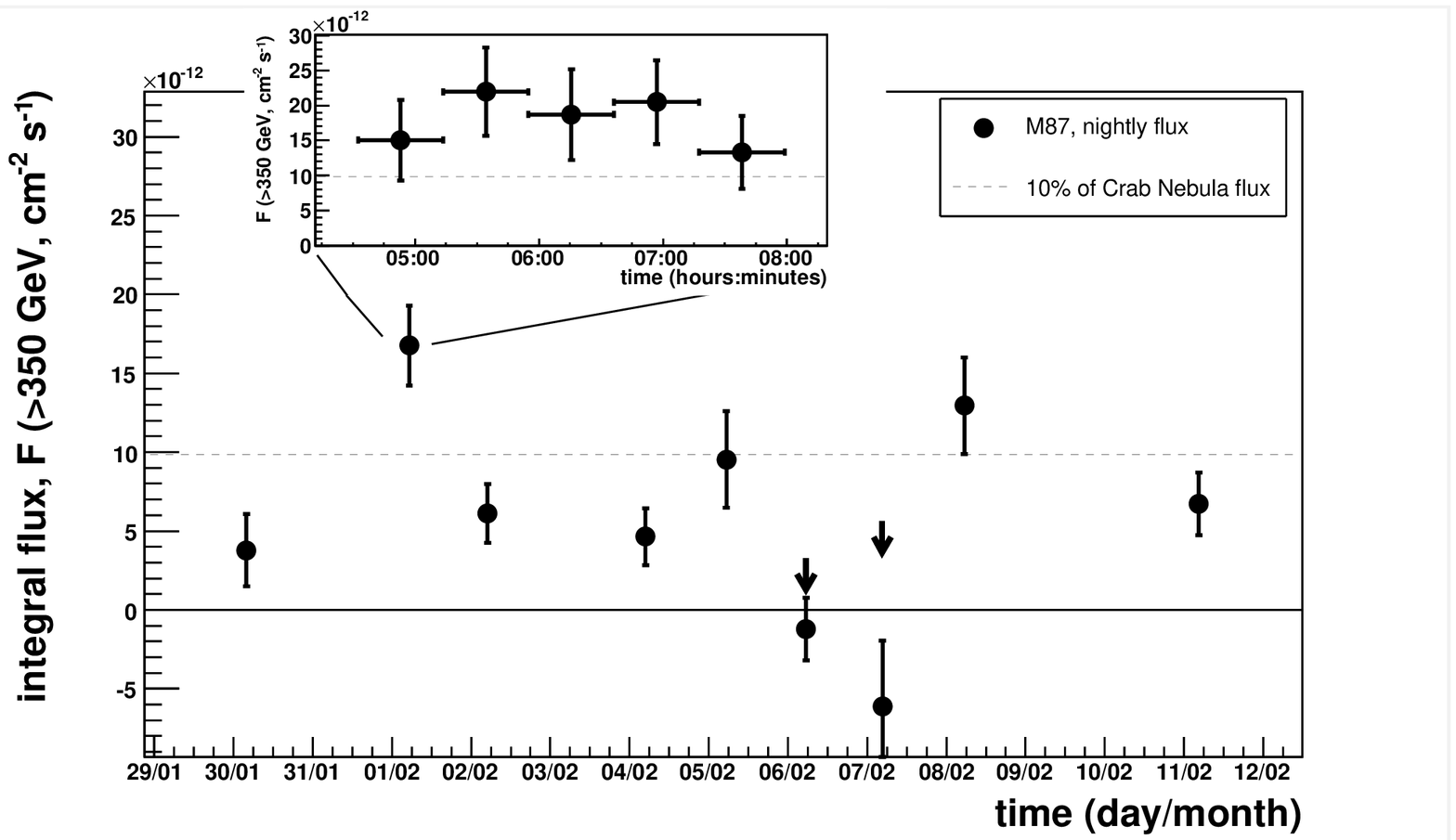}
\caption{\label{fig:lc} The night-by-night light curve for M\,87
as measured from 2008 January 30 (MJD 54495) to 2008 February 11
(MJD 54507). The upper panel shows the flux in the energy bin
$150 - 350$\,GeV, being consistent with a constant emission.
The lower panel shows the integral flux above 350\,GeV;
flux variations are apparent on variability timescales down to 1
day. The inlay of the lower panel shows the light curve above 350
GeV in a 40 min time binning for the night with the highest flux
(2008 February 1).
The vertical arrows represent flux upper limits (95\% c.l.)
for the nights with negative excesses.}
\end{center}
\end{figure}

In searching for time variability, the data set was divided into nine
subsets, one per observing night. In Fig.~\ref{fig:lc} we show both
the light curve above the analysis threshold ($150 - 350$\,GeV) and in
the energy range at which MAGIC has the highest sensitivity for a
variability search ($>$350\,GeV).
The low-energy range shows no significant variability with a
$\chi^2_\mathrm{\nu}$ of 12.6/8 (probability of $P=0.13$) for a constant
fit. Instead, in the energy range above 350\,GeV clear variability is
found. A fit by a constant
has a $\chi^2_\mathrm{\nu}$ of 47.8/8 corresponding to
$P=1.1\cdot10^{-7}$. The correlation coefficient between the two
energy bins is $r = -0.25^{+0.40}_{-0.33}$ (1-$\sigma$ errors),
suggesting that there is no significant correlation, but we note
rather large error bars in the low energy flux bin. We also
investigated a night-by-night variability. There are five pairs of
observations on consecutive nights in the total data set. We
calculated individual probabilities $S_i$ for these pairs to have
the same flux level and the corresponding significances. We then
computed a combined significance $S_\mathrm{comb}$ 
\citep{bityukov}:
$ S_\mathrm{comb} = \left( {\sum{S_i}} \right) / {\sqrt{n}} $,
with $n=5$. 
We interpret the resulting $S_\mathrm{comb} = 5.6\,\sigma$
as a proof that the flux varies on timescales of
1 day or below.
Note that the 1 day variability is claimed from this
combined analysis rather than from the 2008 February 1 flare alone.
We find our statistics not sufficient enough to determine the
flare shape. Given the number of the observed changes in the flux
level, the data belong to a complex of two, if not three,
sub-flares.

We also looked for shorter time variability, but in none of the
observation nights there is a significant flux variation in the
two energy bands. A typical example in a 40 minute binning is
shown in the inset in Fig.~\ref{fig:lc} for 2008 February 1.

\begin{figure}
\begin{center}
\includegraphics[width=0.875\linewidth]{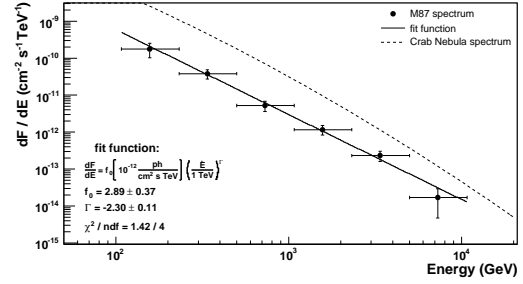}
\caption{The differential energy spectrum of M\,87 for the total
data sample. The horizontal error bars represent width of the energy bins.
The best-fit function, assuming a power law, is given
by the solid curve. The Crab Nebula spectrum \citep{crab} is given
by the dashed curve for reference.} \label{fig:spectot}
\end{center}
\end{figure}

\begin{figure}
\begin{center}
\includegraphics[width=0.875\linewidth]{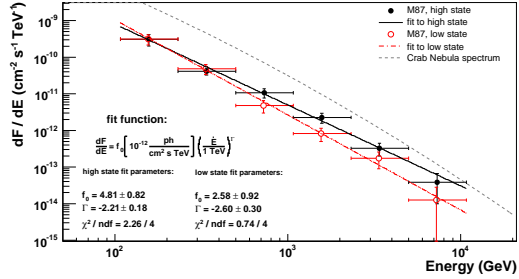}
\caption{Differential energy spectra of M\,87 divided into
high (filled circles) and low (open circles) states. See text for the details.
The best-fit functions, assuming power laws, are
given by the black solid and red dashed-dotted curves,
respectively.}
\label{fig:specHL}
\end{center}
\end{figure}

The averaged differential energy spectrum of M\,87
(Fig.~\ref{fig:spectot}) extends from $\sim 100$~GeV to $\sim
10$~TeV and can be well approximated by a power law:
\[
{\frac{\mathrm{d}F}{\mathrm{d}E}}=(2.89\pm0.37)\times10^{-12}\left({\frac{E}{1\,\mathrm{TeV}}}\right)^{-2.30\pm0.11}\,
\frac 1 {\mbox{TeV}\,\mbox{cm}^{2}\,\mbox{s}}.
\]
The errors are statistical only. We estimate an 11\% systematic
uncertainty in the normalization and 0.20 for the spectral index
\citep{crab}. The measured values are in good agreement with the
H.E.S.S.\ (spectral index $\Gamma = -2.2 \pm 0.15$,
\citealt{HESS-M87}) and VERITAS ($\Gamma = -2.31 \pm 0.17$,
\citealt{VERITAS-M87}) results. The observed spectrum is not
significantly affected by the evolving extragalactic background
light (EBL) due to the proximity of M\,87 \citep{na07}. To
investigate a possible hardening of the spectrum with increasing
absolute flux level, we divided the data sample into \textit{high}
and \textit{low} state subsamples. The \textit{high} sample
comprises the two nights with the highest flux above 350\,GeV
(February 1 and 8), while the \textit{low} state comprises the
nights of lower-flux data (January 30, February 2, 4, and 11).
Both the \textit{high} and \textit{low} state spectra
(Fig.~\ref{fig:specHL}) can be well described by a power law:
\[
{\frac{\mathrm{d}F}{\mathrm{d}E}}= f_0 \left({\frac{E}{1\,\mathrm{TeV}}}\right)^{\Gamma}\,
\left[ \frac {10^{-12}} {\mbox{cm}^{2}\,\mbox{s}\,\mbox{TeV}} \right]
\]
with $f_0^{\mathrm{high}} = (4.81 \pm 0.82)$, $\Gamma^{\mathrm{high}} = (-2.21 \pm 0.18)$
and   $f_0^{\mathrm{low}} = (2.58 \pm 0.92)$, $\Gamma^{\mathrm{low}} = (-2.60 \pm 0.30)$
for the \textit{high} and  \textit{low} states, respectively.
There is a marginal hardening of the spectral index
with the higher flux on the level of $1-2$ standard deviations,
depending on the way the significance is calculated.
This hardening is not significant, which can be a consequence of
the fact that the two flux levels (states) differ by less than a
factor of 2.

%
%

\clearpage

\section{Discussion}
%
%
%

%
M87 is the only non blazar radio galaxy known to emit VHE
$\gamma$-rays and one of the best-studied extragalactic black-hole
systems. To enable long-term studies and assess the variability
timescales of M\,87, the H.E.S.S., VERITAS, and MAGIC
collaborations established a regular, shared monitoring of M\,87
and agreed on mutual alerts in case of a significant detection.
During the MAGIC observations, a strong signal of $8\sigma$
significance was found on 2008 February 1, triggering the other
IACTs as well as {\it Swift} observations. For the first time, we
assessed the energy spectrum below 250 GeV, where our observations
can be well described by a power law that shows no hint of any
flattening.

Our analysis revealed a variable (significance: 5.6\,$\sigma$)
night-to-night $\gamma$-ray flux above 350 GeV, while no
variability was found in the 150--350\,GeV range. We confirm the
$E>730\,\mathrm{GeV}$ short-time variability of M\,87 reported by
\citet{HESS-M87}. 
The observed variability timescale is on the order of or even
below 1 day, restricting the emission region to a size of $R\leq
\Delta t\,c\,\delta = 2.6 \times 10^{15}\,\mathrm{cm} =
2.6\,\delta R_S$. The Doppler factor $\delta$ is only relevant for
an emission region not expanding while traversing the jet. In case
of an expanding-jet hypothesis, the \textit{initial} radius of the
expanding shell is given by $R^* < c\,\Delta t$. The emission can
occur very close to the BH, provided that the ambient photon
density is low enough as to allow the propagation of VHE $\gamma$
rays. Otherwise the emission region must be located farther away
from the BH. In the latter case, the variability constraints can
be met only if the emitting plasma does not substantially expand
while traversing the jet, or if it moves with $\delta \gtrsim 10$.

There exists no lower limit on the size of HST-1, and thus the
flux variability cannot dismiss HST-1 as possible origin of the
TeV flux. During the MAGIC observations, however, HST-1 was at a
historically low X-ray flux level, whereas at the same time the
luminosity of the M\,87 core was at a historical maximum (D.
Harris 2008, private communication). This strongly supports the core of M\,87 as
the VHE $\gamma$-ray emission region.

Our data alone cannot put strong constraints on VHE $\gamma$-ray
emission models. The relatively hard VHE spectrum found for M\,87
($\Gamma\approx -2.3$) is not unique among the extragalactic VHE
$\gamma$-ray sources if one considers intrinsic spectra, i.e.
EBL corrected. Also, we did not measure a high-energy spectral
cut-off. The found marginal spectral hardening may be interpreted
as a similarity to other blazars detected at VHE, where such
hardening has often been observed.

Our results show that a dense TeV monitoring, as exercised by
ground-based IACTs, has revealed highly interesting rapid flares
in M\,87. This fastest variability observed so far in TeV
$\gamma$-rays from M\,87 restricts the size of the 
$\gamma$-emission region to the order of $R_S$ of the central BH of M\,87
and suggests the core of M\,87 rather than HST-1 as the origin of
the TeV $\gamma$-rays. Results from the entire monitoring
campaign, comprising data from other IACTs, will appear in a
separate paper.

\acknowledgements
{We would like to thank the Instituto de Astrofisica de Canarias for the
excellent working conditions at the Observatorio del Roque de los
Muchachos in La Palma. The support of the German BMBF and MPG, the
Italian INFN and Spanish MCI is gratefully acknowledged. This work was
also supported by ETH Research Grant TH 34/043, by the Polish MNiSzW
Grant N N203 390834, and by the YIP of the Helmholtz Gemeinschaft.
We thank Dan Harris for providing preliminary results on {\it Chandra}
observations.

}

\end{document}